\documentclass{article} 
\usepackage[final]{colm2026_conference}

\usepackage{microtype}
\usepackage{hyperref}
\usepackage{url}
\usepackage{longtable}
\usepackage{booktabs}
\usepackage{graphicx}
\usepackage{amsmath}
\usepackage{amssymb}
\usepackage{layout}
\usepackage{multicol}
\usepackage{xcolor} 
\usepackage{booktabs}      
\usepackage{threeparttable} 
\usepackage{tabularx}
\usepackage{subcaption}

\usepackage{lineno}

\usepackage{amsthm}

\theoremstyle{definition}
\newtheorem{definition}{Definition}

\definecolor{darkblue}{rgb}{0, 0, 0.5}
\hypersetup{colorlinks=true, citecolor=darkblue, linkcolor=darkblue, urlcolor=darkblue}

\title{The Shrinking Lifespan of LLMs in Science}

\author{Ana~Trišović \\
Computer Science \& Artificial Intelligence Laboratory \\
Massachusetts Institute of Technology \\
32 Vassar St, Cambridge, MA 02139, USA \\
\texttt{ana\_tris@mit.edu} \\
}

\begin{document}

\ifcolmsubmission
\linenumbers
\fi

\maketitle

\begin{abstract}
Scaling laws describe how language model capabilities grow with compute and data, but say nothing about how long a model \textit{matters} once released. We introduce \textit{time-to-peak} and \textit{lifespan} as measures of model obsolescence and use them to characterize the scientific adoption trajectories of 62 LLMs across more than 108k citing papers (2019-2025), separating active adoption from background citation to recover per-model trajectories that citation counts cannot resolve.
We find that a model's longevity is shaped more by \textit{when} it was released than by its characteristics: release year predicts time-to-peak and lifespan more strongly than architecture, openness, or scale.
LLM adoption follows an inverted-U curve (rising after release, peaking, and then declining), but this pattern is rapidly compressing - a pattern we term the capability treadmill. Each successive release year is associated with a 27\% shorter time-to-peak and a 23\% shorter lifespan ($p < 0.001$), robust to minimum-age thresholds and controls for model size.
These adoption-side dynamics are invisible to scaling laws and suggest that specialization on any single model is a depreciating investment, with costs falling on reproducibility and migration.
\end{abstract}

\section{Introduction}
\label{sec:intro}

Scaling laws have become the governing equations of language model
development: given compute budget $C$, dataset size $D$, and
parameter count $N$, one can predict loss with remarkable
precision~\citep{kaplan2020scaling, hoffmann2022training}. Yet
scaling laws describe how models \textit{improve}. They say
nothing about how long a model \textit{matters}. As the field
produces frontier systems at an accelerating cadence, a parallel
question becomes urgent: what governs the scientific relevance of
a language model after it is released?

Over the past five years,
large language models (LLMs) have crossed a threshold from being objects
of study to serving as \textit{instruments} of study. Researchers
in biomedicine, social science, and the humanities now rely on
LLMs to classify documents, extract entities, synthesize
literatures, simulate social agents, and replace human coders in
experimental pipelines~\citep{grossmann2023ai, ziems2024can,
tornberg2023chatgpt}. Recent large-scale surveys confirm the
breadth of this shift: \citet{liang2024mapping} document a steady
rise in LLM-modified content in scientific papers since 2022,
while \citet{liao2025llms_research_tools} report that over 80\% of researchers have already incorporated LLMs into their
workflows. When a model is embedded this deeply in a scientific workflow, its lifespan ceases to be a product-cycle question and becomes an \textit{infrastructure} question. Deprecation does not merely remove a convenience; it breaks pipelines, severs
reproducibility, and forces costly migration to a successor whose
outputs may not be comparable.

Despite growing awareness that LLM adoption in research is
rising~\citep{liao2025llms_research_tools}, surprisingly little is known about what happens
\textit{after} adoption. Does usage of a given model accumulate
steadily, plateau, or decline? How does the answer change across
model generations? And which properties of a model predict
whether it remains scientifically useful or is rapidly
supplanted? Without answers, the field risks what we term a
\textbf{capability treadmill}: scientists cycle through
successive frontier models faster than cumulative methodological
knowledge can consolidate around any one of them, undermining
reproducibility and raising adoption costs for resource-constrained
disciplines. We treat the treadmill as a motivating concern rather
than a problem this paper resolves: lifecycle measurement alone
cannot settle the reproducibility risks of model succession, but
quantifying the pace of succession provides an empirical basis for
discussions of model documentation, behavioral testing, version
locking, and open-weight alternatives.

In this paper, we provide the first large-scale empirical characterization of the \textit{lifecycle} of language models in scientific research. Drawing on a fine-grained analysis of 108k papers drawn from Semantic Scholar, we track adoption and reference trajectories for {62} language models and estimate the shape, duration, and correlates of their scientific relevance. Throughout, we define scientific relevance in terms of a model's use as part of the methodology that generates a scientific result --- classification, extraction, retrieval, embedding, fine-tuning, or simulation --- and exclude ancillary uses such as text polishing, brainstorming, or general writing assistance. We introduce time-to-peak and lifespan as general measures of model obsolescence, and use them to establish three empirical regularities:

\begin{itemize}
  \item \textbf{Scientific adoption of LLMs follows an inverted-U trajectory.} Usage of individual models rises after release, peaks, and then declines as newer alternatives enter. This pattern holds within every release cohort, establishing a baseline ``lifecycle shape'' for LLMs as research instruments. (\S\ref{sec:ucurve})
  \item \textbf{The LLM adoption arc is compressing.} Successive cohorts of models reach their peak faster and decline sooner, so that the effective window of scientific relevance is \textit{shrinking} over time. Models released before 2020 accumulated adoption over three to four years; post-2022 models peak within one to two.  (\S\ref{sec:compression})
  \item \textbf{Predictors of longevity and use.} Using cross-sectional regressions with release year fixed effects, we identify which model-level attributes (openness of weights, parameter count, architectural class, training type, API access, and provider type) predict longer or shorter time to peak adoption and scientific lifespan. (\S\ref{sec:halflife})
\end{itemize}


Together, these results contribute to an emerging science of
language model ecosystems: where scaling laws characterize the
supply side of model development, we provide a demand-side
complement, showing how model relevance evolves with time and
competitive displacement.



\section{Related Work}
\label{sec:related}

\textbf{Scaling laws and temporal dynamics of LLMs.}
A foundational strand of research characterizes how capabilities scale with compute, data, and parameters~\citep{kaplan2020scaling,hoffmann2022training}, with related work documenting emergent capabilities that arise discontinuously with scale~\citep{wei2022emergent}. These studies ask what a model \textit{can} do. We ask how long it \textit{matters}. A supply-side literature examines how model quality degrades over time: GPT-3.5 and GPT-4 exhibit behavioral drift within three months~\citep{chen2024chatgpt}, model quality degrades in 91\% of model-dataset pairs through patterns not reducible to concept drift~\citep{vela2022temporal}, deprecated API usage persists at 25--38\% across code-generation models~\citep{wang2025llms}, and ``LLM Decay'' (the persistence of outdated facts despite authoritative updates) has been attributed partly to citation density bias in training corpora~\citep{de2025llm,de2022artificial}. Closest in spirit to our lifespan construct, \citet{singh2025shelf} formalize the \emph{shelf life} of fine-tuned LLM judges, finding that judges trained on today's generator models transfer poorly to future ones. We complement this supply-side perspective with demand-side bibliometric evidence, showing that adoption lifecycles follow regularities tied to release timing rather than model-level properties.

\textbf{Diffusion of AI and LLMs in science.}
Bibliometric evidence documents steady growth of AI-related research across all scientific domains~\citep{HAJKOWICZ2023102260}, with near-exponential uptake of foundation models in Linguistics, Computer Science, and Engineering~\citep{bommasani2021opportunities,trivsovic2025rapid,kim2025discovering}.
Adoption has been mapped through keyword analysis~\citep{gao2024quantifying}, broad LLM surveys~\citep{10.1145/3664930,10.3389/frai.2023.1270749}, model-level citation studies revealing uneven uptake in non-CS fields~\citep{pramanick2026transforming}, and platform-level analyses showing that broad, all-purpose AI adoption beyond science is concentrated among a small number of models~\citep{osborne2024ai}.
We frame this uptake through diffusion theory~\citep{rogers2014diffusion}, where cumulative adoption follows an S-curve, implying that the \textit{rate} of adoption per period follows an inverted-U trajectory. Existing work establishes \textit{that} scientists adopt LLMs and \textit{how much}. No prior study has characterized the full adoption arc of a specific model, measured whether that arc is compressing, or identified which model characteristics predict longevity.

\textbf{LLM use in scientific practice.}
AI adoption increases individual scientific impact while narrowing topical breadth~\citep{hao2026artificial}, augments R\&D productivity at the organizational level~\citep{besiroglu2024economic}, and AI-assisted papers show a measurable citation-impact premium despite uneven adoption across fields~\citep{gao2024quantifying}. Survey evidence shows 81\% of researchers report active LLM use, with especially high adoption among junior and non-native English-speaking scholars~\citep{liao2025llms_research_tools}, alongside persistent barriers of compute access and data quality~\citep{van2023ai}. 
LLM-assisted writing is rising across disciplines, peaking in Computer Science at up to 17.5\%~\citep{liang2024mapping}, with evidence of co-evolution between scientific vocabulary and model language~\citep{geng2025human}. At least 13.5\% of 2024 PubMed abstracts involve LLMs, with a writing impact surpassing that of the COVID-19 pandemic~\citep{kobak2025delving}. These studies measure adoption at a point in time or in aggregate. We instead track model-level citation trajectories longitudinally to characterize lifecycle shape, compression, and determinants.

\section{Measuring LLM Adoption and Obsolescence}
\label{sec:measuring}

We characterize LLM adoption trajectories over model \emph{age} (time since release) rather than calendar time, enabling comparisons across individual models and \textit{model cohorts} (models released in the same year). We first introduce time-to-peak and lifespan as general measures of model obsolescence, then describe their operationalization in this paper.

\subsection{Definitions}

Let $u_m(\tau)$ denote the \emph{usage} of model $m$ at age $\tau$,
measured by any consistent adoption signal.

\begin{definition}
    \textbf{Time to peak} ($\tau^*_m$) is the age at which usage is highest,
    \[
    \tau^*_m = \arg\max_{\tau}\, u_m(\tau), \qquad u_m^{\max} = \max_{\tau}\, u_m(\tau),
    \]
    measuring how quickly a model reaches maximal uptake after release.
\end{definition}

\begin{definition}
    \textbf{Lifespan} ($\ell_m$) is the length of time usage stays at or above a
    fraction $\theta \in (0,1)$ of its peak. Let $[\tau^-_m, \tau^+_m]$ be the
    contiguous age interval containing the peak on which $u_m(\tau) \ge \theta\, u_m^{\max}$.
    Lifespan is the width of this interval,
    \[
    \ell_m(\theta) = \tau^+_m - \tau^-_m,
    \]
    measuring the duration of a model's sustained relevance.\footnote{Equivalently,
    the model's \emph{shelf-life}: the term captures the perishability of a model.} Unless otherwise stated we report
    $\ell_m \equiv \ell_m(0.5)$, the time usage stays above half its peak.
\end{definition}

Both constructs are defined relative to each model's own peak, making them invariant
to differences in absolute usage volume across models and comparable across release
cohorts of very different sizes. This ensures that models serving smaller disciplines
are not overshadowed by high-volume models in fields such as computer science, and lets
us study the \emph{shape} and timing of the adoption curve independently of its scale.

\subsection{Operationalization in this study}
\label{sec:operationalization}

We instantiate usage as paper-level adoption: $c_{m,y} \equiv u_m(\tau_{m,y})$ is the number of papers adopting model $m$ in calendar year $y$, where model age is $\tau_{m,y} = y - \rho_m$ and $\rho_m$ is the release year. To make trajectories directly comparable in figures, we plot the normalized
trajectory $\tilde{c}_{m,y} = c_{m,y}/c_m^{\max} \in [0,1]$, where $c_m^{\max} = \max_{y'} c_{m,y'}$, which equals $1$ at the
year of highest observed adoption. 

To reduce sensitivity to year-to-year noise in annual counts, we estimate $\tau^*_m$ and
$\ell_m$ from a quadratic fitted separately to each model's normalized annual trajectory rather than from raw counts.
We classify each trajectory by explicit criteria on the fitted quadratic:
\emph{inverted-U} if the fit is concave with an interior peak inside the observed age range;
\emph{still rising} if the implied time-to-peak falls at or beyond the maximum observed age;
and \emph{no interior peak} if the fit is convex (implied time-to-peak $\leq 0$).
Still-rising models are right-censored --- their peak remains unobserved --- and are
excluded from peak and lifespan estimation. For every model with a confirmed inverted-U,
both $\tau^*_m$ and $\ell_m$ are estimated analytically from the fitted quadratic,
yielding an identical sample ($n = 57$) for the two outcomes: lifespan is computed from
the roots of the fitted quadratic at the half-peak threshold, and is defined only where
the decline below half-peak is analytically recoverable.

Our unit of analysis is the deployable model rather than the abstract model family,
because scientific adoption decisions are made over specific releases with distinct
capabilities, sizes, and release dates; Appendix~\ref{app:families} shows that our
results are robust to collapsing multi-generation lineages to a single representative
model.


\subsection{Data}
\label{sec:data}

\textbf{Model Selection Criteria.}
We draw candidate models from the Epoch AI Index~\citep{EpochNotableModels2024, sevilla2022compute} and manually supplement each entry with model size (total parameters) and availability (e.g., open weights). We apply three inclusion criteria. First, the model must be a transformer-based LLM released from 2017 onward~\citep{vaswani2017attention}, hence we exclude pre-transformer architectures (e.g., LSTMs). Second, the model must have been \textit{publicly accessible} at the time of adoption, whether through open weights, open-source code, or a public API. Internally developed models without a public release (e.g., Chinchilla, Flamingo) are excluded. Third, the model must be a \textit{deployable artifact} rather than a methodological contribution: we exclude attention mechanism variants (e.g., ScatterBrain, Linear Transformer), compression techniques (e.g., SqueezeBERT, MobileBERT), and training infrastructure systems (e.g., Megatron-LM), as scientists cite these as techniques rather than adopting them as research tools.

\textbf{Identifying Scientific Adoption.} We use the Semantic Scholar Academic Graph (S2AG)~\citep{kinney2023semantic, wade2022semantic}
(February 2026 snapshot) to track usage of the selected models, supplemented with full-text extraction from the S2ORC corpus~\citep{lo2019s2orc}. Each LLM is linked to its Semantic Scholar Corpus ID, which allows us to retrieve all citing papers. To extract citation contexts from citing papers, we combine pre-extracted plain text from S2ORC with a custom pipeline that parses paper PDFs using Nougat~\citep{blecher2023nougat}. Our text-extraction pipeline covers preprints, open-access papers, and other publicly available publications. We classify each citation sentence as either \textsc{context} (background reference) or \textsc{adoption} (using the model as a tool, with or without modification) via zero-shot prompting of GPT-4.1-mini~\citep{openai2025gpt41} with a three-sentence context window. This boundary is central to our measurement: a citation counts as adoption only when the model is used as part of the methodology generating a scientific result; background references, motivating mentions, and writing assistance do not qualify. Against a hand-labeled calibration sample, the classifier achieves a sentence-level false positive rate of 6.0\%, macro-F1 of 78.0\%, and recall of 57.6\%. Because trajectories are normalized to each model's own peak, the false positive rate is the most relevant metric for our analysis: it governs distortion of the trajectory's \emph{shape}, whereas imperfect recall attenuates absolute counts without systematically reshaping trajectories. For papers citing a single publication that introduces multiple model variants (e.g., Llama 7B and 70B), we disambiguate the specific variant using Llama-3.1-8B~\citep{dubey2024llama}. A paper is labeled as \emph{adopting} a model if at least one of its citation sentences is classified as adoption. Because papers with more citation sentences are mechanically more likely to contain a misclassified sentence, we apply a Bayesian correction that combines hand-labeled paper-level false positive rates with a binomial model of sentence-level classification errors (Appendix~\ref{sec:falsepositives}). To generalize to the full population of citing publications, we construct inverse-probability weights using the complete S2AG citation graph (Appendix~\ref{sec:weighting}).

\textbf{Sample Summary.} 
Our final sample comprises 62 LLM variants spanning 54 model papers, since a single paper may introduce multiple model sizes (e.g., LLaMA; see Appendix Table~\ref{app:sample}). From these, we retrieve 108,514 citing papers, of which 22,504 are classified as adopters; we apply population weights (Appendix~\ref{sec:weighting}) to estimate adoption counts across the full Semantic Scholar corpus. Each model in the sample meets two additional thresholds: at least 8 observed adoption sentences and at least three years of observed adoption, the minimum required to characterize a trajectory. We exclude the 2018 release cohort due to its small size (2 models, not counted among the 62 in our final sample) and truncate the analysis at the end of 2025, dropping 2026 citations to avoid partial-year artifacts.

\subsection{Statistical Models}
\label{sec:stats}

\textbf{Scientific adoption curve.} We model normalized adoption 
as a quadratic function of relative model age:

\begin{equation}
\tilde{c}_{m,y} = \beta_0 + \beta_1 \tau_{m,y} + \beta_2 \tau_{m,y}^2
+ \varepsilon_{m,y}.
\label{eq:ucurve}
\end{equation}

An inverted-U is indicated by $\hat{\beta}_2 < 0$. We fit this curve in two pooled forms: once across all models, and once separately within each release cohort. For each pooled fit the implied peak age is $\hat{\tau}^\dagger = -\hat{\beta}_1/(2\hat{\beta}_2)$, with a 95\% confidence interval obtained via the delta method. This pooled $\hat{\tau}^\dagger$ summarizes the average cohort trajectory and is distinct from the per-model time-to-peak $\tau^*_m$ defined in Sec.~\ref{sec:measuring}, which we estimate separately below.

\textbf{Lifecycle compression.} To test whether successive model cohorts reach peak adoption faster and lose relevance sooner, we regress log time to peak $\tau^*_m$ and log scientific lifespan $\ell_m$ separately on release year ($\rho_m$):

\begin{equation}
\log(\phi_m) = \alpha_0 + \alpha_1 \rho_m + \varepsilon_m, \quad \phi_m \in \{\tau^*_m,\, \ell_m\},
\label{eq:compression}
\end{equation}

\noindent A negative $\hat{\alpha}_1$ indicates compression: each successive cohort peaks or loses relevance $(1 - e^{\hat{\alpha}_1}) \times 100\%$ sooner than the previous one. The log-linear specification ensures fitted values remain positive and implies proportional rather than constant compression. Because $\tau^*_m$ and $\ell_m$ are only defined for models whose peak has been observed, estimation is restricted to the subset $\mathcal{M}^\dagger = \{m : \text{peak}(m) \text{ observed}\}$ (\S\ref{sec:operationalization}).

\textbf{Predictors of scientific longevity and use.} To identify factors predicting lifecycle duration, we regress lifecycle measures $\phi_m \in \{\tau^*_m, \ell_m\}$ on model characteristics:

\begin{equation}
\phi_m = \delta_0 + \boldsymbol{\delta}^\top \mathbf{x}_m + \sum_{\rho \in \mathcal{R}} \xi_\rho \mathbf{1}[\rho_m = \rho] + \varepsilon_m, \quad m \in \mathcal{M}^\dagger,
\label{eq:predictors}
\end{equation}

\noindent where $\mathbf{x}_m$ includes open model status, size (log parameters), architecture class, origin sector (eg., industry) and training type (eg., fine-tuned). Release year fixed effects $\xi_\rho$ isolate within-cohort variation. To measure absolute adoption volume predictors, we include log adoption counts as an additional outcome variable ($\log_{10}\big(\sum_{y} c_{m,y}\big)$), regressing on the same predictors; this specification uses all models, not the peak-observed restriction. Estimation is implemented in PyFixest~\citep{pyfixest}. All regressions use heteroskedasticity-robust standard errors.

%
\section{The Scientific Adoption Curve}
\label{sec:ucurve}

We characterize the typical adoption trajectory of an LLM in
scientific literature by computing normalized adoption counts
(the number of papers citing a model in a given year, scaled by the
model's peak adoption) and examining how this evolves with relative
model age $\tau_{m,y}$, defined as years elapsed since release. We
formally test whether the aggregate trajectory constitutes an
inverted-U using three complementary approaches.
First, we estimate the quadratic regression
(Equation~\ref{eq:ucurve}) with heteroskedasticity-robust standard
errors (HC1). Both the linear and quadratic terms are highly
significant ($p < 0.001$), with the quadratic coefficient
$\hat{\beta}_{\tau^2} = -0.041$ confirming downward concavity.
The implied peak falls at
$\hat{\tau}^\dagger = 3.58$ years
(95\% CI: $3.30$--$3.87$, delta method), well within the observed
age range $(0.5, 6.5)$.
An $F$-test confirms the quadratic specification significantly
improves over the linear ($p < 0.001$).
Second, following \citet{lind2010or}, we test the joint hypothesis
that the slope is positive at the lower bound
($\tau_{\min} = 0.5$) and negative at the upper bound
($\tau_{\max} = 6.5$). The estimated slopes are $+0.250$ and
$-0.236$, respectively, both individually significant
($p < 0.001$).
Third, following \citet{simonsohn2018two}, we split the sample at
$\hat{\tau}^\dagger$ and estimate separate linear regressions on
each arm. The rising phase ($\tau \leq 3.58$; $n = 213$) yields a
positive slope ($p < 0.001$) and the
falling phase ($\tau > 3.58$; $n = 57$) a negative slope
($p = 0.036$), confirming that both arms
are individually significant. All six conditions for a confirmed
inverted-U are satisfied
(Appendix Table~\ref{tab:utest}).
The pattern is not an artifact of the quadratic functional form:
generalized additive model (GAM) fits recover nearly identical
cohort peak ages, and cubic extensions are not preferred in most
cohorts (Appendix~\ref{app:trajrobust}).


\begin{figure}[htb]
\centering
\includegraphics[]{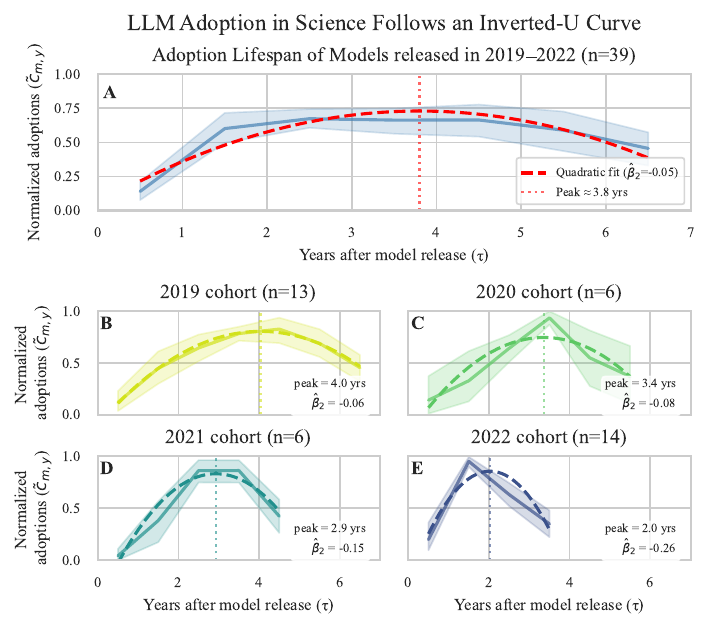}
\caption{\textbf{LLM adoption in science follows an inverted-U
curve that compresses across cohorts.}
\textbf{(A)}~Aggregate trajectory for models released in
2019--2022.
\textbf{(B--E)}~Per-cohort trajectories. Peak shifts earlier each
cohort ($4.0 \to 2.0$ years) and curvature steepens, indicating
lifecycle compression.}\label{fig:ucurve}
\end{figure}

\textbf{Per-cohort trajectories.} The aggregate trajectory in Figure~\ref{fig:ucurve}A is estimated 
on the 2019--2022 cohorts to ensure balanced observation windows.
Figures~\ref{fig:ucurve}B--E decompose the aggregate by release
cohort. The inverted-U holds within every cohort independently, 
with two systematic patterns. First, the peak ($\hat{\tau}^\dagger$) shifts earlier with
each successive cohort: 4.0 years for the 2019 cohort (48 months),
3.4 (2020), 2.9 (2021), 2.0 (2022), and 1.5 (2023, 18 months). Second, the curvature steepens monotonically
($\hat{\beta}_2 = -0.06, -0.08, -0.15, -0.26, -0.45$), indicating that models rise
faster, peak sharper, and decline more steeply with each
generation. Appendix Table~\ref{tab:cohort_quad} reports separate quadratic regressions
by release cohort.

\textbf{Interpretation.}
One natural reading is that the rising phase of the curve reflects
the time required for scientists to discover, evaluate, and
integrate a model into active research workflows, while the falling
phase reflects displacement by newer alternatives that offer
improved capabilities; our data are consistent with this account
but do not identify the mechanism directly. Under this reading, the peak at approximately 43 months (pooled
sample) marks the inflection point where displacement pressure
begins to outweigh continued adoption.



\section{The Compression of the Adoption Arc}
\label{sec:compression}

\begin{figure}[tb]
\centering
\includegraphics[width=0.95\linewidth]{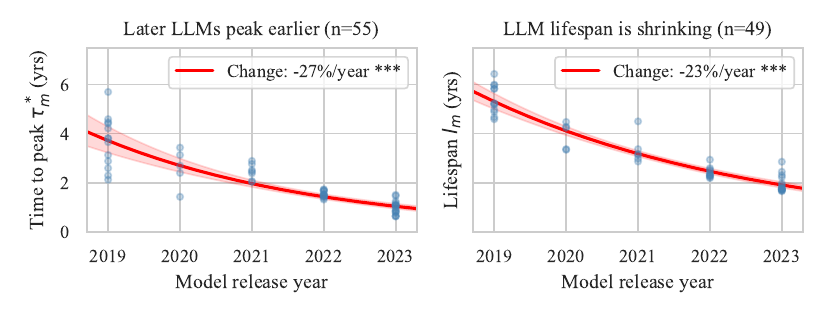}
\caption{Peak adoption age and lifespan per model release year.}
\label{fig:compression}
\end{figure}

Having established that adoption follows an inverted-U on average, we ask whether this arc is \textit{shortening} over time, i.e.,
whether more recently released models reach peak adoption earlier and lose relevance sooner.
Applying the classification criteria of \S\ref{sec:operationalization}, 57 models exhibit a inverted-U and enter the time-to-peak and lifespan estimation; the remainder are convex or still rising and are excluded as right-censored (Appendix Table~\ref{tab:shape_classes} and Figure~\ref{fig:shapes}).
We estimate Equation~(\ref{eq:compression}) separately with time to peak $\tau^*_m$ and lifespan $\ell_m$ as dependent variables, using release year $\rho_m$ as the regressor.
Compression is strong and highly significant: each year of later release
is associated with a $27\%$ reduction in time to peak
($p < 0.001$; $n = 57$).
The decay phase compresses at a comparable rate:
lifespan shrinks by $23\%$ per year of later release
($p < 0.001$; $n = 57$).
Figure~\ref{fig:compression} and Appendix Table~\ref{tab:compression} illustrate these findings.
As an additional robustness check, we vary the minimum observable age required for inclusion: the estimate remains strong and highly significant across window restrictions, attenuating only modestly as the recent cohorts carrying the strongest compression signal drop out --- a censoring artifact would instead collapse it toward zero (Appendix Table~\ref{tab:window} and Figure~\ref{fig:sensitivity}).
We further assess sensitivity of the lifespan estimates to the choice of threshold $\theta$. We observe that the compression rate is stable across the full range tested, varying narrowly between $-22\%$ and $-23\%$ per year with sample size and significance unchanged across all specifications ($n = 49$, $p < 0.001$) (see Appendix Table~\ref{tab:sens_lifespan}). The robustness of both the coefficient magnitude and the sample composition to threshold choice confirms that the lifespan compression finding does not depend on where the half-peak boundary is drawn.
Finally, the estimates are robust to within-family model succession: collapsing each multi-generation lineage (e.g., GPT-2/3/4, LLaMA/Llama~2) to a single representative model leaves the compression coefficients essentially unchanged (Appendix~\ref{app:families}).
%

\textbf{Interpretation.} Models released before 2020 accumulated 
scientific adoption gradually over three to four years. Models released 
after 2022 peak within one to two years and are displaced faster, 
consistent with --- though not direct evidence for --- an
accelerating pace of model development that
shortens the window within which any given model can anchor
scientific practice.
A proxy for model capability is model size. If newer models are genuinely more capable than older ones, faster peak adoption may reflect rational updating by scientists rather than lifecycle compression per se. Controlling for model size ($\log_{10}$ parameters) does not attenuate the compression effect --- time-to-peak compression strengthens to $-30\%$ per year and lifespan compression to $-23\%$ (both $p < 0.001$; Appendix Table~\ref{tab:compression}) --- confirming that compression is a temporal phenomenon driven by cohort turnover rather than a size effect. We also find that larger (i.e., presumably more capable) models tend toward shorter lifespans.

\subsection{What Compresses Faster?}
\label{sec:compression_subgroup}

\begin{figure}
\centering
\includegraphics[width=0.95\linewidth]{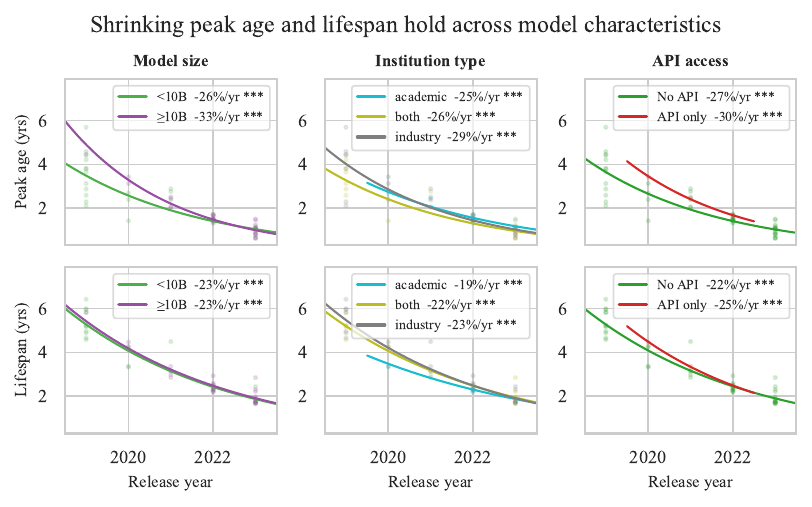}
\caption{Compression rate by model characteristic. }
\label{fig:main_compression_subgroup}
\end{figure}

Compression is nearly universal across subgroups defined by architecture,
training type, parameter count, and institution type
(Figure~\ref{fig:main_compression_subgroup}; Appendix Table~\ref{tab:compression}).
The two exceptions do not reach significance for identifiable reasons rather
than a true absence of compression: encoder-only models (BERT, RoBERTa, and
their variants) are concentrated in a narrow pre-2020 window with limited
release-year variation, and fine-tuned models show a point estimate comparable
to base models on a smaller, more heterogeneous sample ($n = 18$).
The one substantive differential concerns model size: large and small models
lose relevance at the same rate ($-23\%$ per year on lifespan), but larger
models reach their peak faster ($-33\%$ vs.\ $-26\%$ on time to peak, both
$p < 0.001$), suggesting that it is the speed of ascent, not the duration of
relevance, that differs across the parameter scale. Full subgroup estimates
appear in Appendix Table~\ref{tab:compression}.

\section{Predictors of Scientific Longevity and Use}
\label{sec:halflife}

Compression describes a cohort-level trend. We now turn to the complementary model-level question: conditional on release year, which characteristics predict how quickly a model reaches peak adoption, how long it remains actively adopted, and how many total adoptions it accumulates? With Equation~\eqref{eq:predictors}, we find that year fixed effects alone account for the vast majority of variation in lifecycle timing, with model characteristics adding very little explanatory power for lifespan or time to peak. By contrast, adoption volume retains substantially more between-model variation after absorbing year effects, leaving room for model characteristics to contribute (Appendix Tables~\ref{tab:reg_fe_lifespan}, \ref{tab:reg_fe_logAdoptions}, \ref{tab:reg_fe_time_to_peak}).~\footnote{Time to peak and lifespan are estimated for models with a confirmed inverted-U trajectory ($N = 57$). Adoption volume is defined for all models in the sample ($N = 62$).} The clearest model-level results concern adoption volume. Larger models attract significantly more adoptions conditional on release year ($p < 0.05$, attenuating with additional controls), consistent with the outsized role that frontier-scale models play as objects of benchmarking and replication. API-accessible models accumulate more adoptions ($p < 0.10$), offering suggestive evidence that ease of access is a meaningful barrier to scientific uptake independent of model quality or size. Fine-tuned models accumulate fewer adoptions than base-pretrained models ($p < 0.10$), suggesting that task-specific adaptations serve narrower research communities.

For lifecycle shape, results are weaker and largely null once release year is controlled. Larger models tend to reach peak adoption somewhat later ($p < 0.05$, attenuating in the full specification), possibly because their computational demands slow initial diffusion before citations accumulate. Industry affiliation and architecture type show no consistent association with lifespan or time to peak once year effects are absorbed.~\footnote{Release year is a strong and statistically significant predictor of both time to peak and lifespan in all specifications.}

These results are broadly robust to replacing year fixed effects with a binary pre/post-2022 indicator motivated by the structural break following ChatGPT's release (Appendix Tables~\ref{tab:reg_bin_fe_lifespan}, \ref{tab:reg_bin_fe_logAdoptions}, \ref{tab:reg_bin_fe_time_to_peak}). The one notable sensitivity involves model size, which becomes a strong negative predictor of lifespan and time to peak under the coarser control: without granular year effects, size absorbs residual cohort compression, reinforcing the conclusion that release timing rather than model scale drives lifecycle dynamics.

\subsection{What models have longest lifespans?}


To identify the most durable models in our sample, we compute a retention
score: each model's 2025 normalized adoption count as a fraction of its own
historical peak. We restrict to models released before 2022 ($n=26$),
excluding one model (LUKE) whose peak coincides with the final observed year
and is therefore right-censored. 
Appendix Table~\ref{tab:durable} ranks the ten highest-retention pre-2022
models. Every one is a base/pretrained model, spanning all three architecture classes. Architecture and institutional origin show no significant relationship with retention.



\section{Discussion}
\label{sec:discussion}


Our results establish three empirical regularities in the scientific adoption of language models. First, adoption follows an inverted-U, a shape predicted by diffusion theory for any technology facing sequential displacement; the contribution is not the shape itself but its formal confirmation and quantification for LLMs as scientific instruments. Second, the arc is compressing with each generation.
Third, lifecycle duration is governed overwhelmingly by release timing: architecture, size, training type, and openness matter at the margins, but a decoder-only model and an encoder-decoder from the same year have more similar lifecycles than two decoder-only models released three years apart.
%
A model's lifecycle thus appears to be less a function of its intrinsic properties than of the competitive landscape it enters, and---if the release cadence continues to accelerate---architectural innovation alone seems unlikely to insulate a model from compression. For scientists, one possible implication is that the useful life of their chosen instrument is governed not by any property they can evaluate at adoption, but by a market dynamic they cannot observe or control. Isolating the mechanism behind release-time dominance (substitution dynamics versus compositional or measurement shifts) remains an open problem, though capability-side evidence points in a consistent direction: \citet{singh2025shelf} show that fine-tuned LLM judges generalize poorly to responses from newer generator models, one concrete mechanism by which successor releases erode research infrastructure built on their predecessors.
Taken together, these dynamics raise the question of whether model turnover is beginning to outpace the timescales on which scientific methods stabilize. Our data cannot settle this, since we measure model adoption but not the maturation time of scientific practice; we flag it as a direction for future work rather than a conclusion.

The compression of the model lifecycle has at least three consequences worth highlighting. 
First, rapid compression raises the practical cost of staying
current. When each frontier model remains relevant for only one
to two years, researchers face recurring cycles of migration and
re-validation. These costs fall unevenly: well-resourced labs can
absorb them, while researchers in the social sciences,
humanities, or at lower-income institutions face a steeper
burden.
Second, compression threatens reproducibility, though the
severity depends on model openness. For open-weight models, tools
such as Ollama, vLLM, and the Hugging Face ecosystem allow
researchers to run archived weights indefinitely, decoupling
reproducibility from adoption trends. For closed models, no such
safeguard exists: providers can deprecate APIs, alter behavior
through silent updates, or retire a model on commercial timelines
that need not align with scientific ones --- a cost that falls
asymmetrically on work built atop closed infrastructure.
Third, scientific infrastructure that enables cumulative knowledge building  
(genome reference assemblies, versioned databases, validated antibody lots) 
persists long enough for methods to stabilize around it. This raises a 
fundamental question for the field: is the rapid turnover of LLMs compatible 
with the accumulation of reliable, replicable scientific knowledge, or does it 
require new norms around model versioning, archiving, and citation?


More broadly, our results suggest that bibliometric lifecycle analysis can serve as a demand-side complement to scaling laws. Scaling laws tell model developers how much capability they can extract from a given budget. Lifecycle curves tell the research community how long that capability will remain a viable anchor for scientific practice. Together, they define both the production function and the depreciation schedule of language models as scientific instruments. We hope this framing encourages model developers to consider longevity (not just capability at release) as a design objective, and encourages funding agencies and scientific institutions to factor lifecycle risk into their infrastructure planning.

\subsection{Limitations}
\label{sec:limitations}

Several limitations qualify our findings. 
First, our adoption measure relies on Semantic Scholar and S2ORC, which overrepresent English-language, open-access publications. Our inverse-probability weighting partially corrects for this, but adoption patterns in fields with lower open-access rates may differ.
Second, our classification of citation sentences into context versus adoption depends on a zero-shot GPT-4.1-mini classifier. Although we validate against hand-labeled data and apply a Bayesian false-positive correction, residual misclassification noise likely attenuates our estimates, making the patterns we report conservative.
Third, with 62 model variants our sample is small for subgroup regressions, and some null results may reflect limited power rather than genuine absence of effects.
Fourth, we measure scientific adoption through paper-level citations, which capture formal scholarly outputs but miss informal adoption channels. Citation trajectories also reflect processes beyond a model's technical usefulness --- field size, publication growth, citation norms, and network effects --- so we interpret them as measures of adoption rather than intrinsic value. That most models in our sample are general-purpose foundation models reduces (though does not eliminate) the role of field-specific publication cultures. Adoption counts are aggregated annually, smoothing within-year dynamics (see Appendix~\ref{app:monthly}).
Fifth, models that become deeply embedded in standard pipelines may be used without explicit citation, an ambient adoption channel our citation-based measure cannot observe. This undercounting disproportionately shortens the measured lifespans of older, established models, so our compression estimates are, if anything, conservative.
Sixth, models still rising at the end of the observation window are excluded as right-censored, so our lifespan estimates should be read as lower bounds on true population lifespans.
Finally, lifespan is inferred relative to each model's own peak, implicitly treating all models as equally important regardless of adoption volume; an absolute measure such as years sustaining at least $k$ adopting papers could yield different conclusions.
These limitations point to natural directions for future work: characterizing field-level adoption dynamics, moving to monthly-resolution trajectory analysis, and explicitly modeling the competitive dynamics of model succession.




\section*{Ethics Statement}

\textbf{Data sources and privacy.}
This study uses two categories of data: metadata about publicly released language models (parameter counts, release dates, organizational affiliation, and availability) drawn from publicly accessible model cards, technical reports, and documentation; and bibliometric data which includes scientific papers and their metadata (publication year, venue, institution type, and citation counts) drawn from publicly available academic databases. Neither dataset contains personally identifiable information, sensitive personal data, or data collected from human participants. No IRB approval was required.

\textbf{Conflicts of interest.}
The author declares no conflicts of interest. 

\textbf{LLM usage disclosure.} The author used Anthropic's Sonnet 4.6 Extended for identification and implementation of statistical tests for robustness checks, for code review and for exporting regression tables and figures.

\textbf{Dual-use considerations.} Our findings characterize aggregate adoption patterns and do not identify individual researchers or institutions. We note that lifecycle predictions could in principle inform strategic release timing by model providers; however, the same information can help the research community anticipate infrastructure risks and plan accordingly.



\section*{Acknowledgments}

This work is funded by Microsoft, Open Philanthropy/Good Ventures and the Alfred P. Sloan Foundation (G-2025-25164). We acknowledge support from OpenAI Research Credits and the UROP Program at MIT. 
The author acknowledges the MIT SuperCloud and Lincoln Laboratory Supercomputing Center for providing HPC resources that have contributed to the research results reported within this paper.

The author offers special thanks to Neil Thompson for his suggestions and to Alex Fogelson for research assistance. The author thanks Gary King, Hanna Halaburda, Paul T. Scott, Mauricio Tec, Kazimier Smith, Omeed Maghzian, and Parker Whitfill for valuable conversations. The author thanks Emanuele Del Sozzo for proofreading and his support during the data collection process, Joseph Emmens for proofreading and his empirical insights, and Adem Bizid for reviewing LLM data. 
The author thanks the undergraduate students who assisted with data collection: Evan Zhang, Selinna Lin, Denis Siminiuc, Grace Yuan, Emma Li, Sri Saraf, Raul D Campos, Yibo Cheng, Alvin Banh, Dora M. Zhou, Ingrid Tomovski, Kristina Sakayeva, and Maeve Zimmer.

\bibliography{colm2026_conference}
\bibliographystyle{colm2026_conference}

\appendix
\section{Extended Methods}
\label{app:sample}

\subsection{Data Sources}

\textbf{Epoch AI Index.} We draw our initial list of language models from the Epoch AI Index~\citep{EpochNotableModels2024, sevilla2022compute}, filtering to retain only models released as reusable artifacts (excluding purely architectural contributions such as the original Transformer~\citep{vaswani2017attention}). We manually supplement the dataset with model size (total trainable parameters) and availability (downloadable weights, open source software or API-only). For models documented with an associated paper, we link each to its Semantic Scholar Corpus ID.

\textbf{Semantic Scholar Academic Graph.}
We retrieve citing papers from the Semantic Scholar Academic Graph ~\citep{kinney2023semantic, lo2019s2orc}, using the February 2026 snapshot. To maximize text coverage, we combine two extraction pipelines: (1)~pre-extracted plain text from S2ORC~\citep{lo2019s2orc}, which includes inline citation annotations, and (2)~a custom pipeline that downloads paper PDFs and parses them with Nougat~\citep{blecher2023nougat}, a model trained on academic documents that standardizes bibliographies. In-text citations are extracted via regular expressions; when available, we default to the S2ORC text.

\subsection{Identifying model adopters}

We classify each citation sentence into three categories reflecting the depth of model engagement: \textsc{context} (background reference), \textsc{uses} (deploying the model without modification), and \textsc{extends} (fine-tuning or retraining). Classification is performed via zero-shot prompting of GPT-4.1-mini~\citep{openai2025gpt41} using a three-sentence context window, which outperformed other approaches we tested (Table~\ref{tab:classification-prompt}).

\begin{table}[h]
\centering
\small
\begin{tabularx}{\linewidth}{X}
\toprule
\textbf{System Instruction} \\
\midrule
You are an expert in many areas of the scientific literature, with a specialty in how machine learning models are used. \\
\midrule
\textbf{Prompt Template} \\
\midrule
You will pretend to be the author of some sentences from an academic paper which reference \texttt{\{model\_descriptor\}}. Your goal is to determine the extent to which you used the cited model in your paper. The specific citation which references the foundation model is highlighted using HTML style \texttt{<cite>} brackets. Choose from the following list to determine the extent to which you adopted the model: \newline\newline
\begin{tabular}[t]{@{}l@{\hspace{6pt}}p{0.85\linewidth}@{}}
(1) & I merely referenced the cited model as relevant background, for its methodology, or its dataset. \\
(2) & I use the cited model or a part of the model itself, but I don't alter the weights. \\
(3) & I make updates to the cited model's weights, through additional gradient-based training such as fine-tuning. \\
\end{tabular} \newline\newline
Please respond in JSON format only, with key \texttt{"answer"} and value either 1, 2, or 3. \newline\newline
The sentences are as follows: \newline
\texttt{"\{multisentence\}"} \\
\bottomrule
\end{tabularx}
\caption{Prompt template for classifying the depth of foundation model adoption in citing papers. The LLM is asked to role-play as the paper's author and select one of three adoption levels based on the context surrounding a highlighted citation.}
\label{tab:classification-prompt}
\end{table}

\textbf{Model Disambiguation.}
Some papers introduce multiple models under a single Semantic Scholar ID (e.g., Llama 7B and 70B). For citations to such papers, we disambiguate the specific model variant using Llama-3.1-8B~\citep{dubey2024llama}, prompted with the model family name, the list of variants, and the three-sentence citation context. When disambiguation remains unclear, we distribute the citation weight across variants proportionally to the distribution of unambiguous citations (or evenly if all are ambiguous).

\textbf{Paper-Level Aggregation.}
We assign each paper a single label via a hierarchy: \textsc{extends} if any sentence is classified as such, \textsc{uses} if at least one \textsc{uses} sentence exists but no \textsc{extends}, and \textsc{context} otherwise. We define \emph{adoption} as \textsc{uses} or \textsc{extends}.

\subsubsection{Empirical Calibration}
\label{sec:falsepositives}

Papers with more citation sentences have a mechanically higher chance of containing at least one misclassified sentence. Sentence-level false positive rates are straightforward to estimate by manual inspection, but paper-level rates require a model of how sentence errors aggregate.

We correct for this using Bayes' theorem. Let $FP(p)$ denote the event that paper $p$ is a false positive, and let $\hat{e}$ and $\hat{u}$ be the observed counts of \textsc{extends} and \textsc{uses} sentences. For a paper with $n$ citing sentences:

\begin{equation}
\P_n(FP(p)) = \frac{\P_n(FP(p) \mid \hat{e}{=}m, \hat{u}{=}k) \;\cdot\; \P_n(\hat{e}{=}m, \hat{u}{=}k)}{\P_n(\hat{e}{=}m, \hat{u}{=}k \mid FP(p))}
\label{eq:bayes}
\end{equation}

The numerator terms are estimated empirically (by sampling and hand-labeling papers with specific sentence configurations) and observed directly from the data. The denominator is modeled parametrically as described below.

\paragraph{Adoption False Positives ($FP_a$).}
A paper is an adoption false positive if it contains no true \textsc{uses} or \textsc{extends} sentences. We model the distribution of misclassified sentences using a binomial with rate $r_{c \to u} + r_{c \to e}$ (the sum of context-to-uses and context-to-extends error rates from our confusion matrix, 2.8\% and 1.8\% respectively), normalized to condition on at least one error:

\begin{equation}
\P_n(\hat{u}{+}\hat{e} = 1 \mid FP_a(p)) = \frac{\mathrm{Binom}(n,\; r_{c \to u} + r_{c \to e})(1)}{1 - \mathrm{Binom}(n,\; r_{c \to u} + r_{c \to e})(0)}
\end{equation}

For $n \in \{1, 3, 5, 9, 13\}$, we hand-label 20 sentences for each of two configurations: $(\hat{e}, \hat{u}) = (1,0)$ and $(0,1)$, weighting by their frequencies in the dataset to obtain $\P_n(FP_a(p) \mid \hat{e} + \hat{u} = 1)$. Values for intermediate $n$ are linearly interpolated. By design, all parameter choices yield \emph{upper bounds} on false positive rates, ensuring our corrections are conservative.

\paragraph{Adjustments to Counts.}
Let $f(n)$ be the fraction of papers in a given subset with $n$ citation sentences. The overall false positive rate for that subset is $\P(FP_a(p)) = \sum_i f(i) \cdot \P_i(FP_a(p))$. Given total weight $W = W_c + W_{\text{adopt}}$, we adjust:

\begin{equation}
    W_{\text{adopt}} \leftarrow W_{\text{adopt}} - W \cdot \P(FP_a(p))
\end{equation}


We do not directly correct for false negatives as they are negligible by construction. Because the base rate of adoption is low (roughly 0.41\% of citing papers customize models), a 5\% false positive rate among the much larger pool of context citations inflates adoption counts by $\sim$54\%. By contrast, a 5\% false negative rate reduces the true count by only $\sim$5\%. Additionally, a paper-level false negative requires \emph{all} adoption sentences to be missed, whereas a single misclassified sentence suffices for a false positive.

\subsubsection{Sample Weighting}
\label{sec:weighting}

Our text-extraction pipelines cover papers that are available online, preprint and open-access papers. To generalize estimates to the full population of citing publications, we construct inverse-probability weights using the complete S2AG citation graph.

Fix a year $t$ and model $M$. Let $Y^M_t$ be the total number of papers citing $M$ in year $t$ (from S2AG) and $X^M_t$ the number observed in our sample. A fraction of papers lack publication years; let $X^M_0$ denote these, and let $p_t = X^M_t / \sum_{\tau \neq 0} X^M_\tau$ be the year distribution among dated papers. We define weights so that the observed sample, plus a proportional share of undated papers, recovers the population total:

\begin{equation}
    W_t^M = \begin{cases}
\displaystyle \frac{Y^M_t}{X_t^M + p_t X_0^M} & t \neq 0 \\[10pt]
\displaystyle \sum_{\tau \neq 0} p_\tau W_\tau^M & t = 0
\end{cases}
\label{eq:weights}
\end{equation}

We apply analogous smoothing to $Y^M_t$ to handle missing years in S2AG. When aggregating to the paper level (across multiple cited models), we use the maximum sentence-level weight, which provides a lower bound on the true represented count by the union bound.

\subsection{List of Models}
\begin{multicols}{2}
\begin{itemize}
\item ALBERT
\item AraGPT2 Mega
\item BART
\item BLOOM 176B
\item ByT5-XXL
\item CPM-2
\item CamemBERT
\item CodeGen Mono 16.1B
\item CodeT5 Base
\item CodeT5 Large
\item CodeT5+
\item CogVLM
\item DeBERTa
\item FLAN-T5
\item FLAN-UL2
\item GLM-10B
\item GLM-130B
\item GPT 2
\item GPT 3
\item GPT 4
\item GPT-NeoX-20B
\item Guanaco
\item InCoder 
\item InstructBLIP
\item InstructGPT
\item Kosmos-2
\item LLaMA
\item LLaMA 2
\item LLaVA
\item LUKE
\item LongT5
\item MT-DNN
\item MT0
\item MiniGPT4 (Vicuna finetune)
\item NLLB
\item OPT
\item OPT-IML
\item PolyCoder
\item Pythia-12B
\item RoBERTa Large
\item SciBERT
\item StarCoder
\item T0-XXL
\item T5
\item Tk-Instruct
\item Transformer-XL Large
\item WizardCoder 
\item WizardLM
\item XLM
\item XLM-R XXL
\item XLM-RoBERTa
\item XLNet
\item mBART-50
\item mT5-XXL
\end{itemize}

\end{multicols}


\section{Supplemental Figures and Results}

\begin{table}[htb]
\centering
\caption{Formal inverted-U tests (full sample, 2019--2023).}
\label{tab:utest}
\begin{tabular}{llcc}
\toprule
Test & Condition & Estimate & $p$-value \\
\midrule
\multicolumn{4}{l}{\emph{Quadratic regression}} \\
\quad Concavity ($\hat{\beta}_{\tau^2} < 0$)
 & $\hat{\beta}_{\tau^2} = -0.041$ & & $< 0.001$ \\
\quad Peak within range
 & $\hat{\tau}^\dagger = 3.58 \in (0.5, 6.5)$
 & & --- \\
\quad Peak 95\% CI (delta method)
 & $(3.30, 3.87)$
 & & --- \\[4pt]
\multicolumn{4}{l}{\emph{\citet{lind2010or} test}} \\
\quad Slope at $\tau_{\min} = 0.5 > 0$
 & $+0.250$ & $t = 7.97$ & $< 0.001$ \\
\quad Slope at $\tau_{\max} = 6.5 < 0$
 & $-0.236$ & $t = 6.19$ & $< 0.001$ \\[4pt]
\multicolumn{4}{l}{\emph{\citet{simonsohn2018two} two-lines test}} \\
\quad Rising phase ($\tau \leq 3.58$, $n = 213$)
 & $\hat{\beta} = +0.123$ & & $< 0.001$ \\
\quad Falling phase ($\tau > 3.58$, $n = 57$)
 & $\hat{\beta} = -0.101$ & & $0.036$ \\
\midrule
\multicolumn{4}{l}{All six conditions satisfied: \checkmark} \\
\bottomrule
\end{tabular}
\end{table}


\begin{table}[htb]
    \centering
    \caption{Quadratic adoption curve estimates by release cohort. The 2023 cohort has only three observation years, leaving minimal residual degrees of freedom for the quadratic specification; its estimates should be interpreted cautiously. The compression result is robust to excluding this cohort: restricting to models with at least three years of post-release observation yields a comparable time-to-peak coefficient; see Appendix Table~4).}
    \label{tab:cohort_quad}
    \small
    \begin{tabular}{lrrcrrccl}
    \toprule
    Sample & Models & Obs
    & $\hat{\beta}_\tau$ & $\hat{\beta}_{\tau^2}$
    & Peak (yrs) & Peak (mos) & $R^2$ & Window \\
    \midrule
    \multicolumn{9}{l}{\emph{Per-cohort}} \\
    \quad 2019 & 13 &  91 & $0.45^{***}$ & $-0.06^{***}$ & 4.0 & 48.4 & 0.49 & 7\,yr \\
    \quad 2020 &  6 &  36 & $0.56^{***}$ & $-0.08^{***}$ & 3.4 & 40.3 & 0.38 & 6\,yr \\
    \quad 2021 &  6 &  30 & $0.85^{***}$ & $-0.15^{***}$ & 2.9 & 35.1 & 0.70 & 5\,yr \\
    \quad 2022 & 14 &  56 & $1.05^{***}$ & $-0.26^{***}$ & 2.0 & 24.3 & 0.53 & 4\,yr \\
    \quad 2023 & 19 &  57 & $1.38^{***}$ & $-0.45^{***}$ & 1.5 & 18.4 & 0.46 & 3\,yr \\
    \midrule
    \multicolumn{9}{l}{\emph{Pooled}} \\
    \quad 2019--2022 & 39 & 213 & $0.36^{***}$ & $-0.05^{***}$ & 3.8 & 45.6 & 0.28 & --- \\
    \quad 2019--2023 & 58 & 270 & $0.29^{***}$ & $-0.04^{***}$ & 3.6 & 43.0 & 0.17 & --- \\
    \bottomrule
    \multicolumn{9}{l}{\footnotesize
    $^{***}\,p<0.001$. HC1 robust standard errors.
    Peak $= -\hat{\beta}_\tau / (2\hat{\beta}_{\tau^2})$.
    } \\
    \end{tabular}
\end{table}

\begin{table}[htb]
\centering
\caption{Trajectory classification by release cohort.}
\label{tab:shape_classes}
\small
\begin{tabular}{lrrrr}
\toprule
Release year & Inverted-U & No interior peak & Rising & Total \\
\midrule
2019 & 13 & 0 & 0 & 13 \\
2020 &  5 & 0 & 1 &  6 \\
2021 &  7 & 1 & 0 &  8 \\
2022 & 13 & 2 & 1 & 16 \\
2023 & 19 & 1 & 1 & 21 \\
\bottomrule
\end{tabular}
\end{table}

\begin{figure}[h]
    \caption{Adoption trajectories are classified by explicit criteria on the fitted quadratic into inverted-U, still-rising, and no-interior-peak (convex) classes. Classification determines eligibility for time-to-peak and lifespan estimation.}
    \centering
    \includegraphics{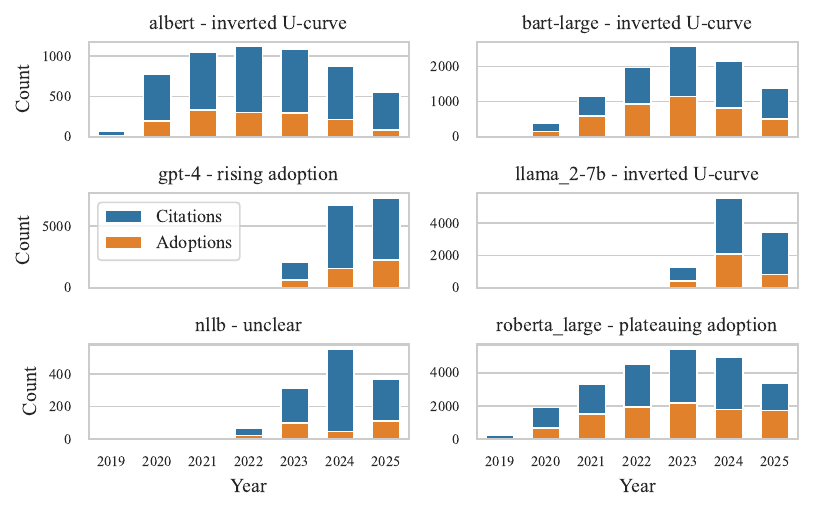}
    \label{fig:shapes}
\end{figure}

\begin{figure}[h]
    \centering
    \begin{subfigure}[b]{\textwidth}
        \centering
        \includegraphics[]{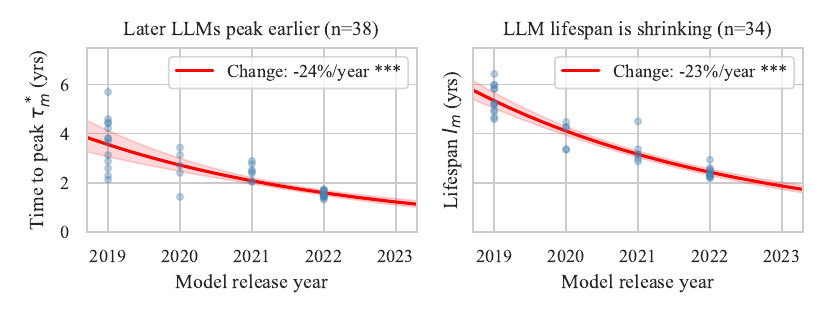}
        \caption{Every model has at least four observed years (2022-2025 or longer).}
        \label{fig:sub1}
    \end{subfigure}
    \vfill
    \begin{subfigure}[b]{\textwidth}
        \centering
        \includegraphics[]{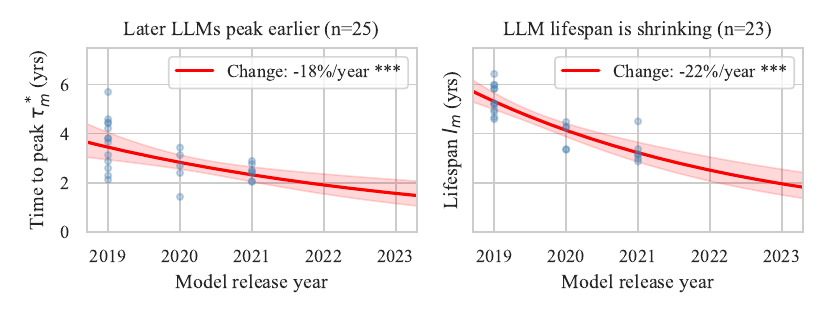}
        \caption{Every model has at least five observed years (2021-2025 or longer).}
        \label{fig:sub2}
    \end{subfigure}
    \caption{Sensitivity analysis.}
    \label{fig:sensitivity}
\end{figure}

\begin{table}[htb]
\centering
\caption{Window-restriction robustness for time-to-peak compression.}
\label{tab:window}
\small
\begin{tabular}{lrlr}
\toprule
Min.\ observable age & $n$ & \% change/yr & $p$ \\
\midrule
$\geq 2$ years & 57 & $-27.2$ ($\hat{\beta} = -0.32$) & $<0.001$ \\
$\geq 3$ years & 38 & $-23.5$ ($\hat{\beta} = -0.27$) & $<0.001$ \\
$\geq 4$ years & 25 & $-17.9$ ($\hat{\beta} = -0.20$) & $<0.001$ \\
\bottomrule
\end{tabular}
\end{table}

\begin{table}[htb]
\centering
\caption{Sensitivity analysis for lifespan, varying threshold $\theta$.}\label{tab:sens_lifespan}
\begin{tabular}{rrrr}
\toprule
$\theta$ & $n$ & $\hat{\beta}$ & \% per year \\
\midrule
0.30 & 49 & $-0.260^{***}$ & $-22.9$ \\
0.40 & 49 & $-0.260^{***}$ & $-22.7$ \\
0.50 & 49 & $-0.260^{***}$ & $-22.6$ \\
0.60 & 49 & $-0.250^{***}$ & $-22.5$ \\
0.70 & 49 & $-0.250^{***}$ & $-22.5$ \\
\bottomrule
\end{tabular}
\par\smallskip
\footnotesize $^{***}p<0.001$.
\end{table}

\subsection{Trajectory Classification Robustness}
\label{app:trajrobust}

To verify that the inverted-U shape and cohort peak ages are not artifacts of the
quadratic functional form, we compare the quadratic fits with generalized additive
models (GAMs) and with cubic extensions. Table~\ref{tab:gam} shows that GAM and
quadratic peak-age estimates agree closely in every cohort.
Table~\ref{tab:cubic} compares quadratic and cubic fits by release cohort. The cubic
term is not preferred for the 2019--2021 cohorts. For the 2022 cohort, the cubic
improvement reflects a sharper, asymmetric peak rather than a departure from the
inverted-U; the GAM fit confirms a single interior maximum. For the 2023 cohort, only
three distinct annual age points are observed, so the quadratic is a saturated
specification and the cubic term is not identified.

\begin{table}[htb]
\centering
\caption{Cohort-level peak-age estimates: GAM versus quadratic fits.}
\label{tab:gam}
\small
\begin{tabular}{lrr}
\toprule
Release year & GAM peak age & Quadratic peak age \\
\midrule
2019 & 4.3 & 4.0 \\
2020 & 3.4 & 3.2 \\
2021 & 3.0 & 2.9 \\
2022 & 1.6 & 2.0 \\
2023 & 1.5 & 1.5 \\
\bottomrule
\end{tabular}
\end{table}

\begin{table}[htb]
\centering
\caption{Quadratic versus cubic fits by release cohort. The 2023 cohort is omitted:
with only three distinct annual age points, the quadratic is a saturated specification
and the cubic term is not identified.}
\label{tab:cubic}
\small
\begin{tabular}{lrrrrc}
\toprule
Release year & $R^2$ quadratic & $R^2$ cubic & $F$ (cubic term) & $p$ & Cubic preferred \\
\midrule
2019 & 0.49 & 0.50 &  0.32 & 0.58  & No \\
2020 & 0.38 & 0.40 &  0.88 & 0.35  & No \\
2021 & 0.39 & 0.44 &  3.15 & 0.08  & No \\
2022 & 0.42 & 0.52 & 12.90 & 0.001 & Yes \\
\bottomrule
\end{tabular}
\end{table}

\subsection{Monthly-Resolution Analysis}
\label{app:monthly}

At monthly resolution, the pooled adoption curve peaks at 37.9 months
(95\% CI: 36.0--39.9), consistent with the annual estimate of 3.58 years.
The two-lines test of \citet{simonsohn2018two} confirms the inverted-U
at this resolution as well: on model-month observations, the rising
phase ($n = 2{,}212$) and falling phase ($n = 822$) are both
individually significant ($p < 0.001$).
We conduct the main analysis at annual resolution because annual aggregation
absorbs publication seasonality and the sparse monthly signal of
low-citation models.

\subsection{Robustness to Model Families}
\label{app:families}

Within-family succession could in principle induce correlated adoption
trajectories, as a successor from the same lineage may displace its
predecessor more directly than an unrelated model. The scope of this
concern is limited in our sample: only a few clearly identifiable
multi-generation lineages exist (e.g., GPT-2/GPT-3/GPT-4, OPT/OPT-IML,
BLOOM/BLOOMZ, LLaMA/Llama~2), and most models have no direct successor
within the observation window. Our main specification also exploits
variation across release cohorts rather than within lineages, which
limits sensitivity to within-family dependence. Conceptually, our unit
of analysis is the deployable model rather than the abstract family,
because scientific adoption decisions are made over specific releases
with distinct capabilities, sizes, and release dates.

As a direct check, we re-estimate the compression regressions after
collapsing each lineage to a single representative model
(Table~\ref{tab:families}). The compression pattern remains significant
and of nearly identical magnitude after collapsing, consistent with the
effect being driven by cross-cohort succession rather than
within-family artifacts: removing within-family duplication leaves the
estimate essentially unchanged.

\begin{table}[htb]
\centering
\caption{Robustness of lifecycle compression to collapsing model
families. Each row reports the release-year coefficient from
Equation~\eqref{eq:compression}; standard errors in parentheses.
Lineages are collapsed to the earliest release; estimates are
essentially unchanged, confirming that compression is driven by
cross-cohort succession rather than within-family duplication.}
\label{tab:families}
\small
\begin{tabular}{llrlrcc}
\toprule
Outcome & Sample & $n$ & $\hat{\beta}$ (log scale) & \% change/yr & 95\% CI & $p$ \\
\midrule
Time to peak & Full             & 57 & $-0.318$ (0.025) & $-27.2$ & $[-30.7,\,-23.5]$ & $<0.001$ \\
Lifespan     & Full             & 57 & $-0.251$ (0.014) & $-22.2$ & $[-24.3,\,-20.0]$ & $<0.001$ \\
Time to peak & Family-collapsed & 47 & $-0.313$ (0.027) & $-26.9$ & $[-30.6,\,-22.9]$ & $<0.001$ \\
Lifespan     & Family-collapsed & 47 & $-0.240$ (0.013) & $-21.3$ & $[-23.4,\,-19.3]$ & $<0.001$ \\
\bottomrule
\end{tabular}
\end{table}

\begin{table}[htbp]
\centering
\caption{Lifecycle compression by release year across model subgroups. The \% change column reports the implied annual proportional change. Subgroups are defined by weight availability (open vs.\ closed), architecture, training type (base vs.\ fine-tuned), parameter count, and institution type.}
\label{tab:compression}
\begin{tabular}{l r r r}
\toprule
 & & \multicolumn{2}{c}{\% change/yr} \\
\cmidrule{3-4}
Group & $n$ & No controls & $+\log_{10}$ params \\
\midrule
\multicolumn{4}{l}{\textit{Panel A: Time to Peak $\tau^*$}} \\
\addlinespace[2pt]
\quad All models          & 55 & $-27\%^{***}$ & $-30\%^{***}$ \\
\quad Open                & 53 & $-27\%^{***}$ & $-30\%^{***}$ \\
\quad No API              & 52 & $-27\%^{***}$ & $-30\%^{***}$ \\
\quad Decoder             & 28 & $-31\%^{***}$ & $-32\%^{***}$ \\
\quad Encoder             & 11 & $-18\%$       & $-20\%$       \\
\quad Encoder-decoder     & 16 & $-30\%^{***}$ & $-30\%^{***}$ \\
\quad Base                & 37 & $-28\%^{***}$ & $-32\%^{***}$ \\
\quad Fine-tuned          & 18 & $-22\%$       & $-23\%$       \\
\quad $<$10B params       & 25 & $-26\%^{***}$ & $-29\%^{***}$ \\
\quad $\geq$10B params    & 30 & $-33\%^{***}$ & $-33\%^{***}$ \\
\quad Academic            &  9 & $-25\%^{***}$ & $-26\%^{***}$ \\
\quad Both                & 17 & $-26\%^{***}$ & $-25\%^{**}$  \\
\quad Industry            & 29 & $-29\%^{***}$ & $-31\%^{***}$ \\
\addlinespace[6pt]
\multicolumn{4}{l}{\textit{Panel B: Scientific Lifespan $\ell$}} \\
\addlinespace[2pt]
\quad All models          & 49 & $-23\%^{***}$ & $-24\%^{***}$ \\
\quad Open                & 47 & $-23\%^{***}$ & $-24\%^{***}$ \\
\quad No API              & 46 & $-23\%^{***}$ & $-24\%^{***}$ \\
\quad Decoder             & 25 & $-24\%^{***}$ & $-24\%^{***}$ \\
\quad Encoder             & 10 & $-16\%$       & $-15\%$       \\
\quad Encoder-decoder     & 14 & $-24\%^{***}$ & $-25\%^{***}$ \\
\quad Base                & 34 & $-23\%^{***}$ & $-24\%^{***}$ \\
\quad Fine-tuned          & 15 & $-19\%^{***}$ & $-20\%^{***}$ \\
\quad $<$10B params       & 22 & $-23\%^{***}$ & $-25\%^{***}$ \\
\quad $\geq$10B params    & 27 & $-23\%^{***}$ & $-23\%^{***}$ \\
\quad Academic            &  6 & $-19\%^{***}$ & $-19\%^{***}$ \\
\quad Both                & 14 & $-22\%^{***}$ & $-27\%^{***}$ \\
\quad Industry            & 29 & $-23\%^{***}$ & $-24\%^{***}$ \\
\bottomrule
\end{tabular}
\begin{tablenotes}\footnotesize
\item \textit{Notes:} Each cell reports the implied annual proportional change from $\hat{\beta}$ in $\ln(\text{DV}) = \alpha + \beta \cdot \text{release\_year} + \varepsilon$, estimated with HC3 robust standard errors. The right column adds $\log_{10}(\text{parameters})$ as a covariate. $n$ reports the number of models in each subgroup. $^{***}p<0.001$; $^{**}p<0.01$; $^{*}p<0.05$.
\end{tablenotes}
\end{table}


\label{app:supplemental}


\begin{table}[ht]
\centering
\begin{threeparttable}
\caption{Predictor analysis. Outcome: \textit{time\_to\_peak}. 
}
\label{tab:reg_fe_time_to_peak}
\begin{tabular}{lccc}
\toprule
 & \textit{M1: Size} & \textit{M2: + Org} & \textit{M3: Full} \\
\midrule
log(Param.) & 0.26** & 0.22* & 0.18 \\
\quad & \textit{(0.10)} & \textit{(0.09)} & \textit{(0.12)} \\
API access &  &  & 0.13 \\
\quad &  &  & \textit{(0.27)} \\
Fine-tuned &  & -0.13 & -0.19 \\
\quad &  & \textit{(0.19)} & \textit{(0.22)} \\
Encoder &  &  & -0.46 \\
\quad &  &  & \textit{(0.59)} \\
Encoder-decoder &  &  & 0.17 \\
\quad &  &  & \textit{(0.33)} \\
Industry &  & -0.08 & -0.09 \\
\quad &  & \textit{(0.12)} & \textit{(0.19)} \\
Both (acad.+ind.) &  & -0.25 & -0.19 \\
\quad &  & \textit{(0.21)} & \textit{(0.23)} \\
\midrule
\multicolumn{4}{l}{\textit{Year fixed effects}} \\
\quad 2020 & -1.28** & -1.32*** & -1.52*** \\
\quad & \textit{(0.39)} & \textit{(0.39)} & \textit{(0.33)} \\
\quad 2021 & -1.47*** & -1.47*** & -1.64*** \\
\quad & \textit{(0.36)} & \textit{(0.37)} & \textit{(0.34)} \\
\quad 2022 & -2.62*** & -2.59*** & -2.77*** \\
\quad & \textit{(0.38)} & \textit{(0.41)} & \textit{(0.41)} \\
\quad 2023 & -3.09*** & -2.99*** & -3.16*** \\
\quad & \textit{(0.36)} & \textit{(0.34)} & \textit{(0.47)} \\
\midrule
N & 55 & 55 & 55 \\
$R^2$ & 0.797 & 0.804 & 0.829 \\
Year FE & \checkmark & \checkmark & \checkmark \\
\bottomrule
\end{tabular}
\begin{tablenotes}
\small
\item $\dagger~p<0.10$, $*~p<0.05$, $**~p<0.01$, $***~p<0.001$. HC1 robust standard errors.
\end{tablenotes}
\end{threeparttable}
\end{table}
\begin{table}[ht]
\centering
\begin{threeparttable}
\caption{Predictor analysis. Outcome: \textit{lifespan}. 
}
\label{tab:reg_fe_lifespan}
\begin{tabular}{lccc}
\toprule
 & \textit{M1: Size} & \textit{M2: + Org} & \textit{M3: Full} \\
\midrule
log(Param.) & 0.21* & 0.19† & 0.16 \\
\quad & \textit{(0.09)} & \textit{(0.10)} & \textit{(0.11)} \\
API access &  &  & -0.03 \\
\quad &  &  & \textit{(0.17)} \\
Fine-tuned &  & -0.13 & -0.09 \\
\quad &  & \textit{(0.14)} & \textit{(0.17)} \\
Encoder &  &  & -0.18 \\
\quad &  &  & \textit{(0.35)} \\
Encoder-decoder &  &  & -0.10 \\
\quad &  &  & \textit{(0.20)} \\
Industry &  & 0.24 & 0.29 \\
\quad &  & \textit{(0.15)} & \textit{(0.19)} \\
Both (acad.+ind.) &  & 0.23 & 0.28 \\
\quad &  & \textit{(0.19)} & \textit{(0.22)} \\
\midrule
\multicolumn{4}{l}{\textit{Year fixed effects}} \\
\quad 2020 & -1.65*** & -1.60*** & -1.60*** \\
\quad & \textit{(0.28)} & \textit{(0.28)} & \textit{(0.31)} \\
\quad 2021 & -2.31*** & -2.24*** & -2.21*** \\
\quad & \textit{(0.35)} & \textit{(0.37)} & \textit{(0.40)} \\
\quad 2022 & -3.40*** & -3.23*** & -3.26*** \\
\quad & \textit{(0.24)} & \textit{(0.29)} & \textit{(0.33)} \\
\quad 2023 & -3.78*** & -3.70*** & -3.76*** \\
\quad & \textit{(0.22)} & \textit{(0.24)} & \textit{(0.32)} \\
\midrule
N & 49 & 49 & 49 \\
$R^2$ & 0.920 & 0.924 & 0.924 \\
Year FE & \checkmark & \checkmark & \checkmark \\
\bottomrule
\end{tabular}
\begin{tablenotes}
\small
\item $\dagger~p<0.10$, $*~p<0.05$, $**~p<0.01$, $***~p<0.001$. HC1 robust standard errors.
\end{tablenotes}
\end{threeparttable}
\end{table}
\begin{table}[ht]
\centering
\begin{threeparttable}
\caption{Predictor analysis. Outcome: \textit{logAdoptions}. 
}
\label{tab:reg_fe_logAdoptions}
\begin{tabular}{lccc}
\toprule
 & \textit{M1: Size} & \textit{M2: + Org} & \textit{M3: Full} \\
\midrule
log(Param.) & 0.31** & 0.28* & 0.21 \\
\quad & \textit{(0.10)} & \textit{(0.11)} & \textit{(0.13)} \\
API access &  &  & 0.70† \\
\quad &  &  & \textit{(0.37)} \\
Fine-tuned &  & -0.29 & -0.36† \\
\quad &  & \textit{(0.18)} & \textit{(0.20)} \\
Encoder &  &  & 0.16 \\
\quad &  &  & \textit{(0.36)} \\
Encoder-decoder &  &  & 0.10 \\
\quad &  &  & \textit{(0.22)} \\
Industry &  & 0.43* & 0.34 \\
\quad &  & \textit{(0.21)} & \textit{(0.22)} \\
Both (acad.+ind.) &  & 0.39 & 0.32 \\
\quad &  & \textit{(0.25)} & \textit{(0.26)} \\
\midrule
\multicolumn{4}{l}{\textit{Year fixed effects}} \\
\quad 2020 & -0.95*** & -0.82*** & -0.88*** \\
\quad & \textit{(0.28)} & \textit{(0.24)} & \textit{(0.24)} \\
\quad 2021 & -1.28*** & -1.14*** & -1.16** \\
\quad & \textit{(0.33)} & \textit{(0.32)} & \textit{(0.37)} \\
\quad 2022 & -1.70*** & -1.37*** & -1.19*** \\
\quad & \textit{(0.25)} & \textit{(0.31)} & \textit{(0.33)} \\
\quad 2023 & -1.21*** & -1.00*** & -0.79* \\
\quad & \textit{(0.26)} & \textit{(0.30)} & \textit{(0.34)} \\
\midrule
N & 62 & 62 & 62 \\
$R^2$ & 0.372 & 0.438 & 0.478 \\
Year FE & \checkmark & \checkmark & \checkmark \\
\bottomrule
\end{tabular}
\begin{tablenotes}
\small
\item $\dagger~p<0.10$, $*~p<0.05$, $**~p<0.01$, $***~p<0.001$. HC1 robust standard errors.
\end{tablenotes}
\end{threeparttable}
\end{table}

\begin{table}[ht]
\centering
\begin{threeparttable}
\caption{Predictor analysis. Outcome: \textit{time\_to\_peak}. 
}
\label{tab:reg_bin_fe_time_to_peak}
\begin{tabular}{lccc}
\toprule
 & \textit{M1: Size} & \textit{M2: + Org} & \textit{M3: Full} \\
\midrule
log(Param.) & -0.40*** & -0.35** & -0.38† \\
\quad & \textit{(0.11)} & \textit{(0.12)} & \textit{(0.22)} \\
API access &  &  & 0.39 \\
\quad &  &  & \textit{(0.47)} \\
Fine-tuned &  & -0.54** & -0.56* \\
\quad &  & \textit{(0.19)} & \textit{(0.23)} \\
Encoder &  &  & 0.01 \\
\quad &  &  & \textit{(0.67)} \\
Encoder-decoder &  &  & 0.10 \\
\quad &  &  & \textit{(0.46)} \\
Industry &  & 0.50† & 0.44 \\
\quad &  & \textit{(0.29)} & \textit{(0.40)} \\
Both (acad.+ind.) &  & 0.27 & 0.23 \\
\quad &  & \textit{(0.32)} & \textit{(0.34)} \\
\midrule
\multicolumn{4}{l}{\textit{Year fixed effects}} \\
\quad Post 2022 & -1.30*** & -1.20*** & -1.12*** \\
\quad & \textit{(0.18)} & \textit{(0.21)} & \textit{(0.30)} \\
\midrule
N & 55 & 55 & 55 \\
$R^2$ & 0.481 & 0.553 & 0.559 \\
Before/After 2022 FE & \checkmark & \checkmark & \checkmark \\
\bottomrule
\end{tabular}
\begin{tablenotes}
\small
\item $\dagger~p<0.10$, $*~p<0.05$, $**~p<0.01$, $***~p<0.001$. HC1 robust standard errors.
\end{tablenotes}
\end{threeparttable}
\end{table}
\begin{table}[ht]
\centering
\begin{threeparttable}
\caption{Predictor analysis. Outcome: \textit{lifespan}. 
}
\label{tab:reg_bin_fe_lifespan}
\begin{tabular}{lccc}
\toprule
 & \textit{M1: Size} & \textit{M2: + Org} & \textit{M3: Full} \\
\midrule
log(Param.) & -0.66*** & -0.55*** & -0.57* \\
\quad & \textit{(0.13)} & \textit{(0.15)} & \textit{(0.24)} \\
API access &  &  & 0.22 \\
\quad &  &  & \textit{(0.51)} \\
Fine-tuned &  & -0.67* & -0.60† \\
\quad &  & \textit{(0.27)} & \textit{(0.31)} \\
Encoder &  &  & 0.04 \\
\quad &  &  & \textit{(0.64)} \\
Encoder-decoder &  &  & -0.19 \\
\quad &  &  & \textit{(0.45)} \\
Industry &  & 0.97** & 0.98* \\
\quad &  & \textit{(0.35)} & \textit{(0.42)} \\
Both (acad.+ind.) &  & 0.95* & 0.91* \\
\quad &  & \textit{(0.41)} & \textit{(0.46)} \\
\midrule
\multicolumn{4}{l}{\textit{Year fixed effects}} \\
\quad Post 2022 & -1.50*** & -1.55*** & -1.54*** \\
\quad & \textit{(0.23)} & \textit{(0.23)} & \textit{(0.32)} \\
\midrule
N & 49 & 49 & 49 \\
$R^2$ & 0.571 & 0.662 & 0.667 \\
Before/After 2022 FE & \checkmark & \checkmark & \checkmark \\
\bottomrule
\end{tabular}
\begin{tablenotes}
\small
\item $\dagger~p<0.10$, $*~p<0.05$, $**~p<0.01$, $***~p<0.001$. HC1 robust standard errors.
\end{tablenotes}
\end{threeparttable}
\end{table}
\begin{table}[ht]
\centering
\begin{threeparttable}
\caption{Predictor analysis. Outcome: \textit{logAdoptions}. 
}
\label{tab:reg_bin_fe_logAdoptions}
\begin{tabular}{lccc}
\toprule
 & \textit{M1: Size} & \textit{M2: + Org} & \textit{M3: Full} \\
\midrule
log(Param.) & -0.08 & -0.01 & -0.03 \\
\quad & \textit{(0.10)} & \textit{(0.10)} & \textit{(0.15)} \\
API access &  &  & 0.78† \\
\quad &  &  & \textit{(0.43)} \\
Fine-tuned &  & -0.42* & -0.40* \\
\quad &  & \textit{(0.19)} & \textit{(0.20)} \\
Encoder &  &  & 0.30 \\
\quad &  &  & \textit{(0.40)} \\
Encoder-decoder &  &  & -0.04 \\
\quad &  &  & \textit{(0.25)} \\
Industry &  & 0.74** & 0.63** \\
\quad &  & \textit{(0.22)} & \textit{(0.23)} \\
Both (acad.+ind.) &  & 0.69* & 0.55* \\
\quad &  & \textit{(0.27)} & \textit{(0.26)} \\
\midrule
\multicolumn{4}{l}{\textit{Year fixed effects}} \\
\quad Post 2022 & 0.04 & 0.02 & 0.14 \\
\quad & \textit{(0.20)} & \textit{(0.22)} & \textit{(0.21)} \\
\midrule
N & 62 & 62 & 62 \\
$R^2$ & 0.012 & 0.220 & 0.300 \\
Before/After 2022 FE & \checkmark & \checkmark & \checkmark \\
\bottomrule
\end{tabular}
\begin{tablenotes}
\small
\item $\dagger~p<0.10$, $*~p<0.05$, $**~p<0.01$, $***~p<0.001$. HC1 robust standard errors.
\end{tablenotes}
\end{threeparttable}
\end{table}


\begin{table}[t]
\centering
\caption{Ten most durable pre-2022 models, ranked by 2025 retention
(adoption in 2025 as a fraction of each model's peak). All ten are
base/pretrained models. LUKE (2020) is excluded as right-censored
(peak at final observed year).}
\label{tab:durable}
\small
\begin{tabular}{llclc}
\toprule
Model & Year & Architecture & Institution & Retention \\
\midrule
RoBERTa Large    & 2019 & Encoder         & Both     & 0.79 \\
GPT-2 (1.5B)     & 2019 & Decoder         & Industry & 0.78 \\
DeBERTa          & 2021 & Encoder         & Industry & 0.72 \\
CodeT5 Base      & 2021 & Encoder-decoder & Both     & 0.67 \\
XLM-RoBERTa      & 2019 & Encoder         & Industry & 0.66 \\
T5-3B            & 2019 & Encoder-decoder & Industry & 0.59 \\
T5-11B           & 2019 & Encoder-decoder & Industry & 0.56 \\
mT5-XXL          & 2020 & Encoder-decoder & Industry & 0.54 \\
SciBERT          & 2019 & Encoder         & Academic & 0.52 \\
Transformer-XL   & 2019 & Decoder         & Both     & 0.47 \\
\bottomrule
\end{tabular}
\end{table}

\end{document}